\documentclass[12pt, reqno]{amsart}

\usepackage{enumerate}
\usepackage{amsmath}
\usepackage{amssymb}
\usepackage{bm}
\usepackage{ascmac}
\usepackage{amsthm}
\usepackage{comment}
\usepackage[margin=33truemm]{geometry}
\usepackage{braket}
\usepackage{tikz}
\usetikzlibrary{calc}

\usepackage{cases}
\usepackage{mathrsfs}

\usepackage{subfiles}

{\theoremstyle{plain}%
  \newtheorem{thm}{\bf Theorem}[section]%
  \newtheorem{lem}[thm]{\bf Lemma}%
  \newtheorem{prop}[thm]{\bf Proposition}%
  \newtheorem{cor}[thm]{\bf Corollary}%
}
{\theoremstyle{remark}
  
}

\newcommand{\dC}{{\mathbb{C}}}%
\newcommand{\MC}[1]{\mathcal{#1}}
\newcommand{\MB}[1]{\mathbb{#1}}
\newcommand{\BM}[1]{{\bm #1}}

\newcommand{\bC}[2]{\begin{bmatrix} #1 \\ #2 \end{bmatrix}}
\newcommand{\MM}[4]{\begin{bmatrix} #1 & #2 \\ #3 & #4 \end{bmatrix}}
\newcommand{\MMM}[9]{\begin{bmatrix}
#1 & #2 & #3 \\ 
#4 & #5 & #6 \\
#7 & #8 & #9 \end{bmatrix}}

\DeclareMathOperator{\rank}{rank}
\DeclareMathOperator{\diag}{diag}
\DeclareMathOperator{\image}{Im}
\DeclareMathOperator{\id}{id}

\makeatletter

\@addtoreset{equation}{section}
\makeatother

\title[A new type of spectral mapping theorem]{A new type of spectral mapping theorem for quantum walks with a moving shift on graphs
}
\author[S. Kubota]{Sho Kubota}
\thanks{S.K. is supported by JSPS KAKENHI (Grant No. 20J01175).}
\author[K. Saito]{Kei Saito}
\author[Y. Yoshie]{Yusuke Yoshie}
\address{Department of Applied Mathematics, Faculty of Engineering, Yokohama National University,
Hodogaya, Yokohama 240-8501, Japan}
\email{kubota-sho-bp@ynu.ac.jp}
\address{Department of Information Systems Creation, Faculty of Engineering, Kanagawa University, Kanagawa, Yokohama 221-8686, Japan}
\email{ft102130ev@jindai.jp}
\address{Department of Mathematics Faculty of Sciences Gakushuin University, Mejiro, Tokyo 171-8588, Japan}
\email{yoshie@math.gakushuin.ac.jp }

\date{}

\begin {document}
\maketitle



\begin{abstract}
The conventional spectral mapping theorem for quantum walks can only be applied for walks employing a shift operator whose square is the identity.
This theorem gives most of the eigenvalues of the time evolution $U$ by lifting the eigenvalues of an induced self-adjoint matrix $T$ onto the unit circle on the complex plane.
We acquire a new spectral mapping theorem for the Grover walk with a shift operator whose cube is the identity on finite graphs.
Moreover, graphs we can consider for a quantum walk with such a shift operator is characterized by a triangulation.  We call these graphs triangulable graphs in this paper.
One of the differences between our spectral mapping theorem and the conventional one is that lifting the eigenvalues of $T-1/2$ onto the unit circle gives most of the eigenvalues of $U$.
\end{abstract}

\section{Introduction}
Quantum walks, which have been researched as quantum versions of random walks \cite{Aharonov, DAharonov, Gudder}, strongly connect to various fields and attract much interest in recent years.
For example, in the field of quantum information, a model of quantum walks can be regarded as a generalization of Grover's search algorithm \cite{Grover, Portugal}.

To clarify various properties of quantum walks, analysis of eigenvalues of the time evolution is significant.
The spectral mapping theorem of quantum walks \cite{FFS1, FFS2, FNSS, HKSS14, MOS, SS19} is a useful tool for analyzing eigenvalues and is a fundamental theorem for connecting quantum and classical systems.
For the finite dimension case, the spectral mapping theorem associates the eigenvalues of two matrices, the time evolution $U$ and a self-adjoint matrix $T$, by lifting the eigenvalues of $T$ on to the unit circle on the complex plane.
The above $T$, called the {\it discriminant operator}, is the matrix induced by the time evolution $U$, whose dimension is smaller than the Hilbert space in which the corresponding quantum walk is defined.
That is, one of the usefulness of the spectral mapping theorems is that we can reduce the dimension of the matrix to analyze eigenvalues.
Notably, it is known that the eigenvalues of $T$ of the Grover walk on graphs with flip-flop shift are the same as that of probability transition matrices of the isotropic random walk.
Thus, one of the excellent features of the spectral mapping theorem is that it allows us to apply well-known results for classical models to quantum walks.
For examples of the applications of the spectral mapping theorem, there are some results of the periodicity \cite{Higuchi13, KSTY, KSTY2, Yoshie, Yoshie2} and the limit averaged measure \cite{FNSS, HSegawa17, Segawa13}.

In 2004, Szegedy \cite{Sze04} found a mathematical connection between a class of quantum walks, called the Szegedy walk, and random walks.
In 2014, Higuchi, Konno, Sato, and Segawa \cite{HKSS14} considered a more generalized model of the Szegedy walks, called the twisted Szegedy walks, and derived a spectral mapping theorem in those models. 
This result generalizes the conventional spectral mapping theorem and succeeds in explicitly giving linearly independent eigenvectors belonging to the $\pm 1$- eigenspaces.
In 2017, Matsue, Ogurisu, and Segawa \cite{MOS} refined Szegedy’s result.
They characterized the eigenspace of the time evolution into two parts.
One part is the eigenspace obtained from the corresponding random walks, and the other part is the eigenspace composed by $\pm 1$ eigenvectors.
Currently, the former is generalized to the eigenspaces obtained from the discriminant $T$ and called {\it the inherited eigenspaces}, while the latter is called {\it the birth eigenspaces}.
In 2019, Segawa, Suzuki \cite{SS19} showed a spectral mapping theorem for quantum walks given a more abstract setting that is not restricted to models on finite graphs.
They showed that not only the eigenvalues but also the spectrum of $U$ are derived from that of $T$.

These previous studies considered quantum walks with a shift operator $S$ called flip-flop shift so that $S^2=I$ where $I$ is the identity matrix.
In contrast, this paper employs a shift operator $S_c$ that satisfies $S_c^3=I$.
Remark that quantum walks with such a shift operator $S_c$ cannot be defined for all finite graphs.
The class of graphs we consider are graphs whose arc sets are decomposed into sets of directed cycles of length $3$.
We call graphs belonging to this class {\it triangulable graphs} in this paper.
Usually, $\sigma(T)$, the set of all eigenvalues of $T$, is in the closed interval $[-1,1]$.
However, for the triangulable graph case, $\sigma(T)$ lies in the interval $[-1/2, 1]$ (Lemma \ref{K60}).

Based on this fact, our spectral mapping theorem (Theorem \ref{KK10}) states that the set of eigenvalues of the time evolution $U$ that belong to the inherited eigenspace is given by lifting $\sigma(T-1/2)$ onto the unit circle on the complex plane.
Furthermore, we show eigenvalues of $-1, -\omega, -\omega^2$ are included in the birth eigenspace and clarify multiplicities of each eigenvalue, where $\omega = e^{\frac{2\pi i}{3}}$.

This paper is organized as follows. 
In section 2, we prepare basic terminologies related to graphs, and review the definition of the Grover walk. 
In section 3, we define our quantum walk and confirm that two discriminant operators we define are equal to that of the Grover walk. Then, as mentioned in the above paragraph, we make sure that the smallest eigenvalue of the discriminant operator is not less than $-1/2$.
In section 4, we prepare for guaranteeing that eigenspaces of the time evolution can be decomposed into the inherited eigenspace and the birth eigenspace. 
In section 5, we reveal the inherited eigenspace of our quantum walk. This is exactly a proof of our spectral mapping theorem. 
In section 6, we investigate the birth eigenspace. Specifically, multiplicities of each eigenvalue of the birth eigenspace are determined. 
In section 7, we consider graphs called double cones as a concrete example. Double cones are a family of triangulable graphs. For these graphs, all linearly independent eigenvectors of the birth eigenspace are also revealed. The last section summarizes results obtained in this paper and discusses future directions.

\section{Preliminaries}
Throughout this paper, graphs we treat are all finite, simple and connected.
Let $G=(V, E)$ be a graph with the vertex set $V$, and the edge set $E$. For an edge $uv \in E(G)$, the arc from $u$ to $v$ is denoted by $(u, v)$. In addition, the {\it origin} and {\it terminus} of $e=(u, v)$ are denoted by $o(e)$ and $t(e)$, respectively.  Furthermore, the {\it inverse arc} of $e=(u,v)$ is denoted by $\bar{e}$, that is, $\bar{e}=(v, u)$. We define $\mathcal{A}=\{ (u, v), (v, u) \mid uv \in E(G) \}$, which is {\it the set of the symmetric arcs} in $G$.
Put the Hilbert spaces $\mathcal{H}=\dC^{\mathcal{A}}$ and $\mathcal{K}=\dC^{V}$. 

First, we review the definition of the Grover walk.
The {\it boundary operator} $d \in \MB{C}^{V \times \MC{A}}$ is defined by
\[ d_{v,a} = \frac{1}{\sqrt{\deg v}} \delta_{v, t(a)}. \]
Note that $dd^{*}=I$, where $d^*$ is the adjoint of $d$.
The {\it flip-flop shift operator} $S \in \MB{C}^{\MC{A} \times \MC{A}}$ is defined by
\[ S_{a,b} = \delta_{a,\bar{b}}. \]
The time evolution of the {\it Grover walk} is given by
\[ U=S(2d^{*}d-I). \]
The {\it discriminant operator} $T \in \MB{C}^{V \times V}$ is defined by
\begin{equation} \label{K5001}
T=dSd^{*}.
\end{equation}

Most eigenvalues of $U$ are given by lifting up the eigenvalues of $T$ to the unit circle on the complex plane.
This fact is known as a spectral mapping theorem of quantum walks:

\begin{thm}[\cite{HKSS14}, Proposition~1]
{\it
Let $G = (V,E)$ be a finite simple connected graph,
let $U$ be the time evolution of the Grover walk,
and let $T$ be the discriminant operator.
Then we have
\[ \sigma(U) = \{ e^{\pm i \cos^{-1} (\lambda)} \mid \lambda \in \sigma(T) \} \cup \{ 1 \}^{M_1} \cup \{-1\}^{M_{-1}}, \]
where
\begin{align*}
M_{1} &= |E| - |V| + 1, \\
M_{-1} &= |E| - |V| + \dim \ker (T+1).
\end{align*}
}
\end{thm}

In contrast to the above, the spectral mapping theorem we have obtained in this study is described below.
For detailed notations, see the later sections.
For detailed proof, see Theorem \ref{main thm}, Corollary \ref{K92}, and Lemma \ref{K113}.
Let $\omega = e^{\frac{2\pi}{3} i}$.

\begin{thm} \label{KK10}
{\it
Let $G = (V,E)$ be a finite simple connected triangulable graph,
let $U_c$ be the time evolution of our quantum walk,
and let $T$ be the discriminant operator, respectively.
Then we have
\[ \sigma(U_c) = \{ e^{\pm i \cos^{-1} (\lambda - \frac{1}{2})} \mid \lambda \in \sigma(T) \setminus \{1\} \} \cup \{ 1 \}^{|V|} \cup \{-1\}^{M_{-1}}
\cup \{ -\omega \}^{M_{-\omega}}
\cup \{ -\omega^2 \}^{M_{-\omega^2}},
\]
where 
\begin{align*}
M_{-1} &= \frac{2}{3}|E| - |V| + \dim \ker \left( T+\frac{1}{2} \right), \\
M_{-\omega} &= \frac{2}{3}|E| - |V| + 1, \\
M_{-\omega^2} &= \frac{2}{3}|E| - |V| + 1.
\end{align*}
}
\end{thm}

\begin{figure}
\begin{center}
\begin{tikzpicture}
[scale = 0.7,
v/.style = {circle, fill = black, inner sep = 0.8mm}]
  \draw[->] (-4.5,0) -- (4.5,0);
  \draw (3.3,-0.3) node {$1$};
  \draw (-3.4,-0.3) node {$-1$};  
  \draw[->] (0,-4.5) -- (0,4.5);
  \draw (0.3, 3.3) node {$i$};
  \draw (-0.2,-0.3) node {$0$};
  \draw (0,0) circle [radius = 3];
  %
  \draw[blue, line width = 1.2pt] (-3, 0.04) -- (3,0.04);
  \node[v] (1) at (1.5, 0) {};
  \node[v] (11) at (1.5, 2.6) {};
  \node[v] (12) at (1.5, -2.6) {};  
  \draw[->, dashed, line width = 1pt, red] (1) -- (11);
  \draw[->, dashed, line width = 1pt, red] (1) -- (12);
  \draw (1.8,-0.3) node {$\lambda$};
  \draw (2.7,3.2) node {$e^{i \cos^{-1}(\lambda)}$};
  \draw (2.9,-2.9) node {$e^{-i \cos^{-1}(\lambda)}$};
  \end{tikzpicture}
$\qquad$
\begin{tikzpicture}
[scale = 0.7,
v/.style = {circle, fill = black, inner sep = 0.8mm}]
  \draw[->] (-4.5,0) -- (4.5,0);
  \draw (3.3,-0.3) node {$1$};
  \draw[->] (0,-4.5) -- (0,4.5);
  \draw (0.3, 3.3) node {$i$};
  \draw (-0.2,-0.3) node {$0$};
  \draw (0,0) circle [radius = 3];
  \draw[blue, line width = 1.2pt] (-1.5, 0.04) -- (3,0.04);
  \draw (-1.7, -0.5) node {$-\frac{1}{2}$};
  \node[v] (1) at (0.9, 0) {};
  \node[v] (2) at (-0.6, 0) {};
  \draw[->, red, line width = 1.2pt] (1) to [bend right = 50] (2);
  \node[v] (21) at (-0.6, 2.94) {};
  \node[v] (22) at (-0.6, -2.94) {};
  \draw[->, dashed, line width = 1pt, red] (2) -- (21);
  \draw[->, dashed, line width = 1pt, red] (2) -- (22);
  \draw (1.2,-0.3) node {$\lambda$};
  \draw (-1.5,3.5) node {$e^{i \cos^{-1}(\lambda - \frac{1}{2})}$};
  \draw (-1.6,-3.5) node {$e^{-i \cos^{-1}(\lambda - \frac{1}{2})}$};
\end{tikzpicture}
\end{center}
\caption{Two spectral mapping theorems}
\label{KK55}
\end{figure}
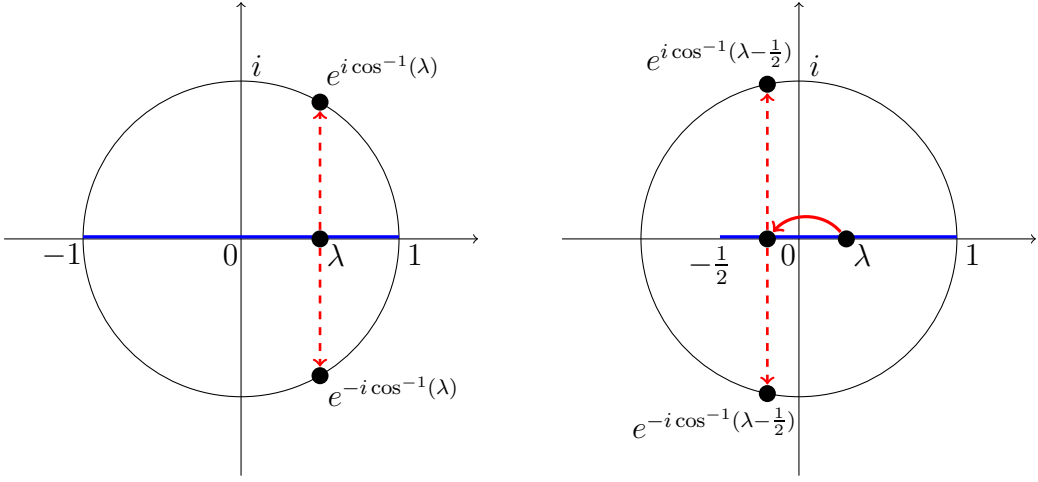

We describe two spectral mapping theorems in Figures~\ref{KK55}.
The left one is the conventional mapping theorem,
and the right one is our theorem in this study.
The blue lines on the real axis show the ranges of the eigenvalues of $T$. 

\section{Quantum walks on triangulable graphs}

\begin{figure}[ht]
\begin{center}
\begin{tikzpicture}
[line width=1pt, scale = 0.8,
v/.style = {circle, fill = black, inner sep = 1.1mm},
g/.style = {circle, fill = black, inner sep = 0mm}]
\node[v] (1) at (2,3.4) {};
\node[v] (2) at (0,0) {};
\node[v] (3) at (4,0) {};
\draw[->] (1) to  (2);
\draw[->] (2) to  (3);
\draw[->] (3) to  (1);
\node[g, label =  above : $o(a)$] (11) at (2,3.4) {};
\node[g, label =  left : $t(a)$] (12) at (0,0) {};
\node[g, label =  right : $z$] (13) at (4,0) {};
\node[g, label =  north west : $a$] (21) at (1,1.7) {};
\node[g, label =  below : $\tau(a)$] (22) at (2,0) {};
\end{tikzpicture}
\caption{The next arc $\tau(a)$ in a directed triangle}
\label{K50}
\end{center}
\end{figure}
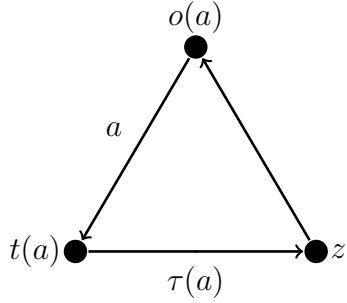

Let $G = (V, E)$ be a graph, and let $\MC{A}$ be the set of symmetric arcs induced by $E$.
For arcs $a_1, a_2, a_3 \in \MC{A}$, the triple $\{ a_1, a_2, a_3 \}$ is said to be a {\it directed triangle}
if $t(a_1) = o(a_2)$, $t(a_2) = o(a_3)$, and $t(a_3) = o(a_1)$.
For a directed triangle $C$ and an arc $a \in C$,
there exists only one vertex $z$ in $C$ that is neither $o(a)$ nor $t(a)$.
Define the {\it next arc} of $a$ by $(t(a), z)$, and we write it as $\tau(a)$.
Figure~\ref{K50} shows the next arc of an arc $a$.
We regard the symbol $\tau$ as a mapping from $\MC{A}$ to $\MC{A}$.
The mapping $\tau$ has the inverse mapping since $\tau \circ \tau \circ \tau = \id_{\MC{A}}$.

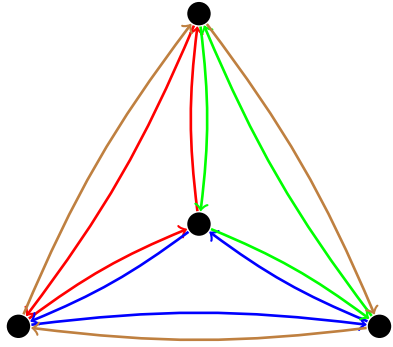
\begin{figure}[ht]
\begin{center}
\begin{tikzpicture}
[line width=1pt, scale = 0.8,
v/.style = {circle, fill = black, inner sep = 1.1mm}]
\node[v] (1) at (3,5.2) {};
\node[v] (2) at (0,0) {};
\node[v] (3) at (6,0) {};
\node[v] (4) at (3,1.7) {};
\draw[->, red] (1) to [bend left = 7] (2);
\draw[->, red] (2) to [bend left = 7] (4);
\draw[->, red] (4) to [bend left = 7] (1);
\draw[->, blue] (2) to [bend left = 7] (3);
\draw[->, blue] (3) to [bend left = 7] (4);
\draw[->, blue] (4) to [bend left = 7] (2);
\draw[->, green] (1) to [bend left = 7] (4);
\draw[->, green] (4) to [bend left = 7] (3);
\draw[->, green] (3) to [bend left = 7] (1);
\draw[->, brown] (1) to [bend left = 7] (3);
\draw[->, brown] (3) to [bend left = 7] (2);
\draw[->, brown] (2) to [bend left = 7] (1);
\end{tikzpicture}
\caption{The complete graph $K_4$ is triangulable}
\label{K21}
\end{center}
\end{figure}

Let $X$ be a finite set,
and let $C_1, C_2, \dots, C_t$ be non-empty subsets of $X$.
We say that $\pi = \{ C_1, C_2, \dots, C_t \}$ is a {\it partition} of $X$ if $X = \bigcup_{i=1}^t C_i$
and $C_i \cap C_j = \emptyset$ for any distinct $i,j \in \{1,\dots,t\}$.
A graph $G$ is {\it triangulable}
if there exists a partition $\pi$ of $\MC{A}$ consisting of directed triangles.
The easiest triangulable graph is the complete graph $K_4$ on $4$ vertices.
It is shown in Figure~\ref{K21}, and the twelve arcs are decomposed into the four colored directed triangles.

Let $G$ be a triangulable graph and
let $\pi$ be a partition of $\MC{A}$ consisting of directed triangles.
For a vertex $x$ of $G$,
we define
\begin{align*}
\pi(x) &= \{ C \in \pi \mid \exists a \in C \text{ s.t. $x = t(a)$} \}, \\
\MC{A}(x) &= \{ a \in \MC{A} \mid t(a) = x \}.
\end{align*}
For a directed triangle $C \in \pi(x)$, there exists only one arc $a \in \MC{A}(x)$.
Conversely, for an arc $a \in \MC{A}(x)$, there exists only one directed triangle $C \in \pi(x)$.
Thus, we have the following:

\begin{lem} \label{K70}
{\it
With the above notation, 
there exists a bijection between $\pi(x)$ and $\MC{A}(x)$.
In particular, $|\pi(x)| = \deg x$.
}
\end{lem}

We define a new quantum walk for triangulable graphs.
Let $G$ be a triangulable graph.
We give another shift operator $S_{c} \in \MB{C}^{\MC{A} \times \MC{A}}$ defined by
\begin{eqnarray*}
(S_c)_{a,b} = \delta_{a, \tau(b)}.
\end{eqnarray*}
Note that $S^{3}_{c}=I$. 
Our quantum walk is defined based on the Grover walk.
Define the time evolution $U_c$ of the quantum walk as
\begin{equation} \label{K5000}
U_{c}=S_{c}(2d^{*}d-I).
\end{equation}
Remak that a partition $\pi$ of $\MC{A}$ consisting of directed triangles is not always unique,
so our quantum walk is defined by specifying a graph $G$ and a partition $\pi$ of $\MC{A}$.

\begin{lem} \label{K01}
{\it
Let $G$ be a graph, and let $T$ be the discriminant operator of $G$ defined in the above.
Then, we have
\[ T = D^{-\frac{1}{2}}AD^{-\frac{1}{2}}, \]
where $A \in \MB{C}^{V \times V}$ is the adjacency matrix of $G$
and $D \in \MB{C}^{V \times V}$ is the degree matrix of $G$,
which are defined by
\[ A_{x,y} = \begin{cases}
1 &\quad \text{if $x \sim y$,} \\
0 &\quad \text{otherwise.}
\end{cases} \]
and $D_{x,y} = (\deg x) \delta_{x,y}$, respectively.
}
\end{lem}

\begin{proof}
We compute the $(x, y)$-component straightforwardly.
  \begin{align*}
  T_{x,y} &= (dSd^*)_{x,y} \\
  &= \sum_{a,b \in \MC{A}} d_{x,a} S_{a,b} d_{y,b} \\
  &= \sum_{a, b \in \MC{A}} \frac{1}{\sqrt{\deg x}} \frac{1}{\sqrt{\deg y}} \delta_{x,t(a)} \delta_{a,\bar{b}} \delta_{y, t(b)} \\
  &= \sum_{a \in \MC{A}} \frac{1}{\sqrt{\deg x}} \frac{1}{\sqrt{\deg y}} \delta_{x,t(a)} \delta_{y, t(\bar{a})} \\
  &= \sum_{a \in \MC{A}} \frac{1}{\sqrt{\deg x}} \frac{1}{\sqrt{\deg y}} \delta_{x,t(a)} \delta_{y, o(a)} \\
  &= \frac{1}{\sqrt{\deg x}} \frac{1}{\sqrt{\deg y}} A_{x,y} \\
  &= (D^{-\frac{1}{2}}AD^{-\frac{1}{2}})_{x,y},
  \end{align*}
so we have the statement.
\end{proof}

Note that the $(x,y)$-component of $T$ is
\begin{align}
T_{x,y} &= \frac{1}{\sqrt{\deg x}} \frac{1}{\sqrt{\deg y}} A_{x,y} \label{KK00} \\
&= \begin{cases}
\frac{1}{\sqrt{\deg x}\sqrt{\deg y}} \qquad &\text{if $x \sim y$,} \\
0 \qquad &\text{otherwise.}
\end{cases} \label{K5003}
\end{align}
As detailed in Subsection~\ref{K201},
the discriminant operator $T$ is similar to a probability transition matrix of the isotropic random walk on $G$.

We give two kinds of discriminant operators defined by $T_{1}=dS_{c}d^{*},  T_{2}=dS^{2}_{c}d^{*}$.
We will show that both operators are actually equal to $T$ of the Grover walk.

\begin{lem} \label{discriminant}
We have $T = T_{1} = T_{2}$.
\end{lem}

\begin{proof}
First, we show that $T_1 = T$.
For $x,y \in V$, we have
\begin{align*}
(T_1)_{x,y} &= (d S_c d^*)_{x,y} \\
&= \sum_{a,b \in \MC{A}} d_{x,a} (S_c)_{a,b} (d^*)_{b,y} \\
&= \sum_{a,b \in \MC{A}} \frac{1}{\sqrt{\deg x}} \frac{1}{\sqrt{\deg y}} \delta_{x, t(a)} \delta_{a, \tau(b)} \delta_{y,t(b)} \\
&= \frac{1}{\sqrt{\deg x}} \frac{1}{\sqrt{\deg y}} \sum_{a \in \MC{A}} \delta_{x, t(a)} \delta_{y,o(a)} \\
&= \frac{1}{\sqrt{\deg x}} \frac{1}{\sqrt{\deg y}} A_{x,y} \\
&= T_{x,y}. \tag{by (\ref{KK00})}
\end{align*}

Next, we show that $T_2 = T$.
Since $S_c^2 = S_c^{-1}$,
we have
\begin{align*}
(T_2)_{x,y} &= (d S_c^{-1} d^*)_{x,y} \\
&= \sum_{a,b \in \MC{A}} d_{x,a} (S_c^{-1})_{a,b} (d^*)_{b,y} \\
&= \sum_{a,b \in \MC{A}} \frac{1}{\sqrt{\deg x}} \frac{1}{\sqrt{\deg y}} \delta_{x, t(a)} \delta_{a, \tau^{-1}(b)} \delta_{y,t(b)} \\
&= \frac{1}{\sqrt{\deg x}} \frac{1}{\sqrt{\deg y}} \sum_{b \in \MC{A}} \delta_{x, o(b)} \delta_{y,t(b)} \\
&= \frac{1}{\sqrt{\deg x}} \frac{1}{\sqrt{\deg y}} A_{x,y} \\
&= T_{x,y}. \tag{by (\ref{KK00})}
\end{align*}
\end{proof}


We give another matrix that is important in analyzing our quantum walk.
Let $G$ be a triangulable graph,
and let $\pi$ be a partition of $\MC{A}$ consisting of directed triangles.
Define the matrix $R \in \MB{C}^{V \times \pi}$ by
\begin{equation} \label{K200}
R_{x, C} = \begin{cases}
1 \qquad &\text{if $C \in \pi(x)$,} \\
0 \qquad &\text{otherwise.}
\end{cases}
\end{equation}

With these matrices,
we find the following 
for the smallest eigenvalue of $T$ of triangulable graphs.

\begin{lem} \label{K60}
{\it
Let $G$ be a triangulable graph,
and let $T$ be the discriminant operator defined as in above.
Then the smallest eigenvalue of $T$ is at least $-1/2$.
}
\end{lem}

\begin{proof}
Let $A$ be the adjacency matrix of $G$,
and let $D$ be the the degree matrix of $G$.
Then,  we first have
\begin{equation}
RR^{\top} = 2A + D. \label{K210}
\end{equation}
Indeed,
 \begin{align*}
(RR^{\top})_{xy} &= \sum_{C \in \pi} R_{x,C} R_{y,C} \\
&= |\pi(x) \cap \pi(y)| \\
&= \begin{cases}
2 \qquad &\text{if $x \sim y$}, \\
\deg x &\text{if $x=y$}, \\
0 \qquad &\text{otherwise.}
\end{cases} \tag{by Lemma~\ref{K70}}
\end{align*}
From Lemma~\ref{K01} and (\ref{K210}), we have
\begin{equation} \label{K90}
T = D^{-\frac{1}{2}}AD^{-\frac{1}{2}} = D^{-\frac{1}{2}} \left \{ \frac{1}{2}(RR^{\top} - D)  \right\} D^{-\frac{1}{2}}
= \frac{1}{2}(D^{-\frac{1}{2}}R)(D^{-\frac{1}{2}}R)^{\top} - \frac{1}{2}I.
\end{equation}
Since the matrix $(D^{-\frac{1}{2}}R)(D^{-\frac{1}{2}}R)^\top$ is positive-semidifinite,
the smallest eigenvalue of $T$ is at least $-1/2$.
\end{proof}

\section{Decomposition of eigenspaces}
In analyzing the spectrum of $U_c$,
we will decompose the Hilbert space $\MC{H}$ in which our quantum walks are defined into two subspaces,
called the {\it inherited eigenspace} and the {\it birth eigenspace},
and try to construct eigenvectors in each space.
However, even if eigenvectors in each subspaceb are completely constructed,
basis consisting of eigenvectors of the total space is not always revealed in general.
To clarify basis of the total space, we have to choose appropriate subspaces of $\MC{H}$.
In this section, we consider conditions on such appropriate subspace.

We denote by $0$ the zero vector for any Hilbert space.
For a matrix $M \in \MB{C}^{n \times n}$,
we write $\ker(M - \lambda I)$ as $\ker(M - \lambda)$,
omitting the identity matrix.
Let $M \in \MB{C}^{n \times n}$ be a square matrix of size $n$,
and let $W$ be a subspace of the vector space $\MB{C}^n$.
We define the multiset $\sigma(M|_W)$ by
\[ \sigma(M|_W) = \{ \lambda \in \MB{C} \mid \ker(M - \lambda) \cap W \neq \{0\} \} \]
with the multiplicity of $\lambda \in \sigma(M|_W)$ equal to $\dim \ker(M - \lambda) \cap W$.
We call this multiset $\sigma(M|_W)$ the {\it spectrum of $M$ restricted to $W$}.
The multiset $\sigma(M|_{\MB{C}^n})$ is nothing but the conventional spectrum $\sigma(M)$ of the matrix $M$.
Clearly,
\[ \ker(M - \lambda) \supset (\ker(M - \lambda) \cap W) \oplus (\ker(M - \lambda) \cap W^{\perp}), \]
but reverse inclusion does not hold in general.

\begin{lem} \label{K04}
{\it
With the above notation,
if $MW \subset W$ and $MW^{\perp} \subset W^{\perp}$, then we have
\[ \ker(M - \lambda) = (\ker(M - \lambda) \cap W) \oplus (\ker(M - \lambda) \cap W^{\perp}) \]
for any $\lambda \in \MB{C}$.
}
\end{lem}

\begin{proof}
It is sufficient to show that the left hand side is included in the right hand side.
Suppose $f \in \ker(M - \lambda)$.
Then there exist $f_1 \in W$ and $f_2 \in W^{\perp}$ such that $f = f_1 + f_2$, so
\[ Mf_1 + Mf_2 = Mf = \lambda f = \lambda f_1 + \lambda f_2. \]
Since $MW \subset W$ and $MW^{\perp} \subset W^{\perp}$,
we have $Mf_1 \in W$ and $Mf_2 \in W^{\perp}$.
Thus, we have $Mf_1 = \lambda f_1$ and $Mf_2 = \lambda f_2$.
Therefore, $f = f_1 + f_2 \in (\ker(M - \lambda) \cap W) \oplus (\ker(M - \lambda) \cap W^{\perp})$.
\end{proof}

Therefore, it is sufficient to assert $MW \subset W$ and $MW^{\perp} \subset W^{\perp}$
in order to guarantee $\sigma(M) = \sigma(M|_W) \cup \sigma(M|_{W^{\perp}})$.
The next lemma supplements the condition $MW^{\perp} \subset W^{\perp}$.

\begin{lem} \label{K03}
{\it
Let $M \in \MB{C}^{n \times n}$ be a unitary matrix, and let $W \subset \MB{C}^{n}$ be a subspace.
Suppose $MW \subset W$.
Then, $MW = W$ if and only if $MW^{\perp} \subset W^{\perp}$.
}
\end{lem}

\begin{proof}
Assume that $MW = W$.
Then $M^{-1}W = W$.
For any $Mf \in MW^{\perp}$ and $g \in W$,
we have $(Mf, g) = (f, M^*g) = (f, M^{-1}g) = 0$ since $f \in W^{\perp}$ and $M^{-1}g \in W$.
Conversely, we assume that $MW^{\perp} \subset W^{\perp}$.
For any $f \in W$, writing $f = MM^{-1}f$,
there exist $f_1 \in W$ and $f_2 \in W^{\perp}$ such that $M^{-1}f = f_1 + f_2$,
so we have $f = Mf_1 + Mf_2$.
Since $f, Mf_1 \in W$ and $Mf_2 \in W^{\perp}$ by the assumption,
it holds that $Mf_2 = 0$.
Thus, we have $f = Mf_1 \in MW$.
\end{proof}

\section{The inherited eigenspace} \label{K30}
Define the matrix $L \in \MB{C}^{\mathcal{A} \times V^3}$ by
$L = \begin{bmatrix} d^* & S_c d^* & S_c^2 d^* \end{bmatrix}$.
Put $\mathcal{L}=\mathrm{Im}L$.
We call the subspace $\MC{L}$ the inherited eigenspace,
and the orthogonal complement $\MC{L}^{\perp}$ the birth eigenspace.
In this section, we analyze the inherited eigenspace of $U_{c}$.
For readers who want to know the result in advance, see Theorem~\ref{main thm}.
Remark that $\mathcal{L}^{\perp} = \ker{d} \cap \ker{dS_{c}} \cap \ker{dS^{2}_{c}}$.
If there is no danger of confusion,
the subscripts of the zero and the identity operator are omitted,
and we simply write as $O$ or $I$.

In order to state the validity of decomposition of $\ker(U_c - \Lambda)$,
we first claim that $\MC{L}$ is invariant under $U_c$.
Define the matrix $\tilde{T} \in \mathbb{C}^{V^3 \times V^3}$ by
\[ \tilde{T}= \MMM{O}{O}{-I}{I}{2T}{2T}{O}{-I}{O}. \]
This matrix is invertible.
Indeed, the inverse matrix is given by
\[ \MMM{2T}{I}{2T}{O}{O}{-I}{-I}{O}{O}. \]

\begin{lem} \label{UL=LT}
We have $U_{c}L=L\tilde{T}$.
\end{lem}
\begin{proof}
Indeed,
\begin{align*}
U_c L &= S_c(2d^*d - I) \begin{bmatrix} d^* & S_c d^* & S_c^2 d^* \end{bmatrix} \\
&= \begin{bmatrix} S_c d^* & 2S_c d^* d S_c d^* - S_c^2 d^*& 2 S_c d^* d S_c^2 d^* - d^* \end{bmatrix} \\
&= \begin{bmatrix} S_c d^* & 2S_c d^* T - S_c^2 d^*& 2 S_c d^* T - d^* \end{bmatrix} \tag{by Lemma~\ref{discriminant}} \\
&= \begin{bmatrix} d^* & S_c d^* & S_c^2 d^* \end{bmatrix} \MMM{O}{O}{-I}{I}{2T}{2T}{O}{-I}{O} \\
&= L \tilde{T},
\end{align*}
which completes the proof.
%
\end{proof}

Since the matrix $\tilde{T}$ is invertible, $\tilde{T}\MC{K}^3 = \MC{K}^3$ holds.
By Lemma~\ref{UL=LT}, we have
\[ U_c \MC{L} = U_c L \MC{K}^3 = L \tilde{T} \MC{K}^3 = L \MC{K}^3 = \MC{L}. \]
%
%
Thus, Lemma~\ref{K04} and Lemma~\ref{K03} derive
$\sigma(U_c) = \sigma(U_c|_{\MC{L}}) \cup \sigma(U_c|_{\MC{L}^{\perp}})$.


Eigenspaces we should clarify are $\ker(U_c - \Lambda) \cap \MC{L}$ for $\Lambda \in \MB{C}$.
The following lemma rewrites $\ker(U_c - \Lambda) \cap \MC{L}$ into kernels of polynomials of $\tilde{T}$.

\begin{lem}
We have $\ker{L}=\ker{(\tilde{T}^{3}+I)}$.
\label{LkerT}
\end{lem}
\begin{proof}
Let
\[ B= \MMM{\frac{1}{2}I}{O}{O}{T}{\frac{1}{2}I}{T}{T}{O}{\frac{1}{2}I}. \]
This is an invertible matrix as the inverse is given by
\[ \MMM{2I}{O}{O}{8T^{2}-4T}{2I}{-4T}{-4T}{O}{2I}. \]
In addition, we have
\begin{equation} \label{BT}
B(\tilde{T}^{3}+I) = \MMM{I}{T}{T}{T}{I}{T}{T}{T}{I}.
\end{equation}
Let $\BM{f} = [f,g,h]^{\top} \in \MC{K}^3$.
The following is immediately seen:
\begin{align*}
\BM{f} \in \ker L 
&\Longleftrightarrow d^*f + S_c d^*g + S_c^2 d^*h = 0 \\
&\Longrightarrow \begin{cases}
d(d^*f + S_c d^*g + S_c^2 d^*h) &= 0, \\
dS_c(d^*f + S_c d^*g + S_c^2 d^*h) &= 0, \\
dS_c^2(d^*f + S_c d^*g + S_c^2 d^*h) &= 0
\end{cases} \\
&\Longleftrightarrow \begin{bmatrix} 0 \\ 0 \\ 0  \end{bmatrix}
= \MMM{I}{T}{T}{T}{I}{T}{T}{T}{I} \begin{bmatrix} f \\ g \\ h  \end{bmatrix} = B(\tilde{T} + I) \BM{f} \\
&\Longleftrightarrow \BM{f} \in \ker (\tilde{T}^3 + I).
\end{align*}
The reverse direction on the second line is confirmed as follows.
We first have
\[ d^*f + S_c d^*g + S_c^2 d^*h \in \ker{d} \cap \ker{dS_{c}} \cap \ker{dS^{2}_{c}} = \MC{L}^{\perp}. \]
On the other hand,
\[ d^*f + S_c d^*g + S_c^2 d^*h = L \BM{f} \in \MC{L}, \]
so it holds that $d^*f + S_c d^*g + S_c^2 d^*h = 0$.
Therefore, we have $\ker{L}=\ker{(\tilde{T}^{3}+I)}$.
\end{proof}

From the above lemma, we can see that
\begin{equation} \label{K212}
\ker(U_c - \Lambda) \cap \MC{L}
= L (\ker(\tilde{T}^3 + 1)(\tilde{T} - \Lambda)).
\end{equation}
for $\Lambda \in \MB{C}$.
Indeed,
\begin{align*}
\ker(U_c - \Lambda) \cap \MC{L}
&= \{ L \BM{f} \in \MC{L} \mid U_cL \BM{f} = \Lambda L \BM{f}, \BM{f} \in \MC{K}^3 \} \\
&= L\{ \BM{f} \in \MC{K}^3 \mid L\tilde{T} \BM{f} = \Lambda L \BM{f} \} \tag{by Lemma~\ref{UL=LT}} \\
&= L\{ \BM{f} \in \MC{K}^3 \mid (\tilde{T} - \Lambda) \BM{f} \in \ker L \} \\
&= L\{ \BM{f} \in \MC{K}^3 \mid (\tilde{T} - \Lambda) \BM{f} \in \ker (\tilde{T}^3 + 1) \} \tag{by Lemma~\ref{LkerT}} \\
&= L (\ker(\tilde{T}^3 + 1)(\tilde{T} - \Lambda)).
\end{align*}
Therefore, consideration of the inherited eigenspace is reduced to
that of the subspace $\ker(\tilde{T}^3 + 1)(\tilde{T} - \Lambda)$.
In the next subsection, we investigate kernels of polynomials of $\tilde{T}$.

\subsection{Kernels of polynomials of $\tilde{T}$}

First, we state that kernels of polynomials of $\tilde{T}$ are decomposed into the sum of several subspaces.

\begin{lem} \label{K20}
{\it
Let $f \in \MB{C}[x]$ be a monic polynomial.
Suppose factorization of $f$ over $\MB{C}$ is
\[ f(x) = (x - \lambda_1)^{m_1}(x - \lambda_2)^{m_2} \cdots (x - \lambda_s)^{m_s}. \]
Let $A \in \MB{C}^{n \times n}$ be a square matrix.
If the sum of subspaces $\ker (A-\lambda_1)^{m_1} + \cdots + \ker (A-\lambda_s)^{m_s}$ is the direct sum, i.e.,
\begin{equation} \label{K211}
\dim (\ker (A-\lambda_1)^{m_1} + \cdots + \ker (A-\lambda_s)^{m_s})
= \sum_{i=1}^s \dim \ker (A-\lambda_i)^{m_i}
\end{equation}
holds, then we have
\[ \ker f(A) = \bigoplus_{i=1}^s \ker(A-\lambda_i)^{m_i}. \]
}
\end{lem}

\begin{proof}
It is clear that
\begin{equation} \label{K11}
\ker f(A) \supset \bigoplus_{i=1}^s \ker(A-\lambda_i)^{m_i},
\end{equation}
so we will show the equality by verifying that the dimensions of both sides are equal.
Remark that
\begin{equation} \label{K10}
\dim \ker (MN) \leq \dim \ker M + \dim \ker N
\end{equation}
holds (see Lemma~3.4.2 in \cite{Ar} for example) for square matrices $M$ and $N$ of same size.
We have
\begin{align*}
\dim \ker f(A) &= \dim \ker \left \{ \Pi_{i=1}^s (A - \lambda_i)^{m_i} \right \} \\
&\leq \sum_{i=1}^s \dim \ker (A-\lambda_i)^{m_i}  \tag{by (\ref{K10})} \\
&= \dim \left \{ \bigoplus_{i=1}^s \ker (A-\lambda_i)^{m_i} \right \}  \tag{by (\ref{K211})} \\
&\leq \dim \ker f(A). \tag{by (\ref{K11})}
\end{align*}
Therefore,
\[ \dim \ker f(A) = \dim \left \{ \bigoplus_{i=1}^s \ker (A-\lambda_i)^{m_i} \right \} \]
holds.
\end{proof}

\begin{cor} \label{K402}
{\it
Let $\omega=e^{\frac{2\pi}{3}i}$.
We have
\begin{enumerate}[(i)]
\item $\ker (\tilde{T}^3 + 1) = \ker (\tilde{T} + 1) \oplus \ker (\tilde{T} + \omega) \oplus \ker (\tilde{T} + \omega^2)$;
\item If $\Lambda \not\in \{ -1, -\omega, -\omega^2 \}$,
\[ \ker (\tilde{T}^3 + 1)(\tilde{T} - \Lambda) = \ker (\tilde{T} + 1) \oplus \ker (\tilde{T} + \omega) \oplus \ker (\tilde{T} + \omega^2) \oplus \ker (\tilde{T} - \Lambda); \]
\item If $\Lambda \in \{ -1, -\omega, -\omega^2 \}$,
\[ \ker (\tilde{T}^3 + 1)(\tilde{T} - \Lambda) = \ker (\tilde{T} - \Lambda)^2 \oplus \ker (\tilde{T} - \omega \Lambda) \oplus \ker (\tilde{T} - \omega^2 \Lambda). \]
\end{enumerate}
}
\end{cor}

\begin{proof}
To apply Lemma~\ref{K20},
we confirm that the sum of subspaces is the direct sum.
This is clear for (i) and (ii),
so we give a proof only for (iii), i.e.,
we show that the sum of subspaces
\[ \ker (\tilde{T} - \Lambda)^2 + \ker (\tilde{T} - \omega \Lambda) + \ker (\tilde{T} - \omega^2 \Lambda) \]
is the direct sum.
It is clear that $\ker (\tilde{T} - \omega \Lambda) \cap \ker (\tilde{T} - \omega^2 \Lambda) = \{0\}$,
so we next confirm $\ker (\tilde{T} - \Lambda)^2 \cap \{ \ker (\tilde{T} - \omega \Lambda) + \ker (\tilde{T} - \omega^2 \Lambda) \} = \{0\}$.
Take $\BM{f} \in \ker (\tilde{T} - \Lambda)^2 \cap \{ \ker (\tilde{T} - \omega \Lambda) + \ker (\tilde{T} - \omega^2 \Lambda) \}$,
and there exist $\BM{f}_i \in \ker (\tilde{T} - \omega^i \Lambda)$ such that $\BM{f} = \BM{f}_1 + \BM{f}_2$.
Then we have
\[ 0 = (\tilde{T} - \Lambda)^2 \BM{f} = (\tilde{T} - \Lambda)^2 \BM{f}_1 + (\tilde{T} - \Lambda)^2 \BM{f}_2
= (\omega \Lambda - \Lambda)^2 \BM{f}_1 + (\omega^2 \Lambda - \Lambda)^2 \BM{f}_2. \]
Since $(\omega \Lambda - \Lambda)^2, (\omega^2 \Lambda - \Lambda)^2 \neq 0$,
it hold $\BM{f}_1 = \BM{f}_2 = 0$.
Thus, we have $\BM{f} = 0$.
\end{proof}

From the right hand side of (\ref{K212}) and Corollary~\ref{K402},
subspaces we should clarify are generalized eigenspaces of $\tilde{T}$.
For these, we prepare several lemmas.

\begin{lem} \label{K301}
Let $\Lambda \in \MB{C}$ with $|\Lambda| = 1$.
If $\Lambda \neq 1$, we have
\[
\ker{(\tilde{T}-\Lambda)} =
\left\{ \begin{bmatrix} f \\ \Lambda^{2}f \\ -\Lambda f \end{bmatrix} \in \MC{K}^3 \; \middle| \;
f \in \ker{\left(T-\frac{\Lambda + \Lambda^{-1} + 1}{2} \right)} \right\}. \\
\]
If $\Lambda = 1$, we have
\[
\ker{(\tilde{T} - 1)} = \left\{ \begin{bmatrix} f \\ f \\ -f \end{bmatrix} \in \MC{K}^3 \; \middle| \; f \in \MC{K} \right\}.
\]
\end{lem}

\begin{proof}
Let $\BM{f}=[f, g, h]^{\top} \in \ker{(\tilde{T}-\Lambda)}$. Then,
\[ \begin{bmatrix} 0 \\ 0 \\ 0 \end{bmatrix} = (\tilde{T}-\Lambda)\begin{bmatrix} f \\ g \\ h \end{bmatrix}
= \begin{bmatrix} -\Lambda f-h \\ f+(2T-\Lambda)g+2Th \\ -g-\Lambda h \end{bmatrix}, \]
so we have $h=-\Lambda f$, $g = \Lambda^{2}f$.
Inserting them to the second component in the above vector, we have
\begin{equation} \label{K300}
(\Lambda -1)\{\Lambda^{2} -\Lambda(2T-1)+1\}f = 0.
\end{equation}
If $\Lambda \neq 1$, then
\[ f \in \ker{(\Lambda^{2}-\Lambda (2T-1)+1)} = \ker{\left(T-\frac{\Lambda + \Lambda^{-1} + 1}{2} \right)} \]
since $\Lambda \neq 0$.
If $\Lambda = 1$,
Equality~(\ref{K300}) holds for arbitrary $f \in \mathcal{K}$.
Thus, we have $\BM{f} \in \left\{ [f, f, -f]^{\top} \in \MC{K}^3 \; \middle| \; f \in \MC{K} \right\}$.
\end{proof}

\begin{lem} \label{K403}
We have 
\[ \ker{(\tilde{T} + \omega^j )^{2}} = \ker{(\tilde{T} + \omega^j)} \]
for $j \in \{1,2\}$.
\end{lem}

\begin{proof}
It is enough to show that $\ker{(\tilde{T} + \omega^j )^{2}} \subset \ker{(\tilde{T} + \omega^j)}$.
Let $\BM{f} = [f,g,h]^{\top} \in \ker{(\tilde{T} + \omega^j )^{2}}$.
Since $(\tilde{T} + \omega^j ) \BM{f} \in \ker(\tilde{T} + \omega^j )$,
there exists $\xi \in \ker(T - 1)$ such that
\begin{equation} \label{K401}
\begin{bmatrix} \xi \\ \omega^{2j} \xi \\ \omega^j \xi \end{bmatrix}
= (\tilde{T} + \omega^j ) \BM{f}
= \begin{bmatrix} \omega^j f - h \\ f+(2T + \omega^j)g + 2Th \\ -g + \omega^j h \end{bmatrix}
\end{equation}
by Lemma~\ref{K301}.
From the first and third components,
we have $h = \omega^j f - \xi$ and $g = \omega^{2j} f - 2 \omega^j \xi$.
Inserting them to the second component, 
we obtain
\begin{equation} \label{K400}
(1-T)f = \frac{\omega^j - 1}{2} \xi.
\end{equation}
Multiplying both sides of this equality by $1-T$,
we have $(1-T)^2 f = \frac{\omega^j - 1}{2} (1-T) \xi = 0$,
so $f \in \ker(1-T)^2$.
However, $\ker(1-T)^2 = \ker(1-T)$ since $T$ is diagonalizable.
Thus, $\frac{\omega^j - 1}{2} \xi = 0$ from Equality~(\ref{K400}).
Since $\xi = 0$, we have $\BM{f} \in \ker (\tilde{T} + \omega^j)$ from Equality~(\ref{K401}).
\end{proof}

\begin{lem} \label{K404}
We have
\[ \ker{(\tilde{T} + 1)^{2}} = \ker(\tilde{T} + 1) \oplus
\left\{ [-\rho, \rho, 0]^{\top} \in \MC{K}^3 \; \middle| \; \rho \in \ker \left( T+\frac{1}{2} \right) \right\}. \]
\end{lem}

\begin{proof}
Let $[f, g, h]^{\top} \in \ker{(\tilde{T}+1)^{2}}$.
Then we have
\[
\begin{bmatrix} 0 \\ 0 \\ 0 \end{bmatrix} = 
(\tilde{T} + 1)^{2} \begin{bmatrix} f \\ g \\ h \end{bmatrix} =
\MMM{1}{1}{-2}{2T+2}{4T^2+2T+1}{4T^2 + 4T - 1}{-1}{-2T-2}{-2T+1} \begin{bmatrix} f \\ g \\ h \end{bmatrix}
\]
By the Gaussian elimination, the above equality becomes
\[ \MMM{1}{1}{-2}{O}{O}{T+\frac{1}{2}}{O}{T+\frac{1}{2}}{O} \begin{bmatrix} f \\ g \\ h \end{bmatrix}
= \begin{bmatrix} 0 \\ 0 \\ 0 \end{bmatrix}. \]
Thus, we have $g, h \in \ker{(T+ \frac{1}{2})}$ and $f=-g+2h$,
so
\begin{align*}
\ker(\tilde{T} + 1)^{2} 
&= \left\{ \begin{bmatrix} -g+2h \\ g \\ h \end{bmatrix} \in \MC{K}^3 \; \middle| \; g,
h \in \ker \left( T+\frac{1}{2} \right) \right\} \\
&= \left\{ \begin{bmatrix} h \\ h \\ h \end{bmatrix} + \begin{bmatrix} -(g-h) \\ g-h \\ 0 \end{bmatrix} \in \MC{K}^3
\; \middle| \; g,
h \in \ker \left( T+\frac{1}{2} \right) \right\} \\
&= \ker(\tilde{T} + 1) +
\left\{ \begin{bmatrix} -\rho \\ \rho \\ 0 \end{bmatrix} \in \MC{K}^3 \; \middle| \; \rho \in \ker \left( T+\frac{1}{2} \right) \right\}. \tag{by Lemma~\ref{K301}}
\end{align*}
Focusing on the third coordinate, it is clear that the sum is the direct sum.
\end{proof}

The following is for determining the dimension of the inherited eigenspace.

\begin{lem} \label{K411}
Let $\Theta_{\lambda}=\cos^{-1}\left(\lambda-\frac{1}{2} \right)$ for $\lambda \in \MB{C}$.
\begin{enumerate}[(i)]
\item For $f \in \ker(T - \lambda)$ with $-\frac{1}{2} < \lambda < 1$, if $L [f, e^{\pm 2i \Theta_{\lambda}}f, -e^{\pm i \Theta_{\lambda}}f]^{\top} = 0$,
then $f = 0$.
\item For $f \in \MC{K}$,
if $L[f,f,-f]^{\top} = 0$, then $f=0$.
\item For $f \in \ker (T + \frac{1}{2})$,
if $L[-f, f, 0] = 0$, then $f = 0$.
\end{enumerate}
\end{lem}

\begin{proof}
(i) Since $-\frac{1}{2} < \lambda < 1$, we have $\frac{\pi}{3} < \Theta_{\lambda} < \pi$.
By Lemma~\ref{LkerT},
\[ \begin{bmatrix} f \\ e^{\pm 2i \Theta_{\lambda}}f \\ -e^{\pm i \Theta_{\lambda}} f \end{bmatrix}
\in \ker L = \ker(\tilde{T}^3+1), \]
so
\[ \begin{bmatrix} 0 \\ 0 \\ 0 \end{bmatrix}
= (\tilde{T}^3+1) \begin{bmatrix} f \\ e^{\pm 2i \Theta_{\lambda}}f \\ -e^{\pm i \Theta_{\lambda}} f \end{bmatrix}
= \MMM{2}{2T}{2T}{*}{*}{*}{*}{*}{*} \begin{bmatrix} f \\ e^{\pm 2i \Theta_{\lambda}}f \\ -e^{\pm i \Theta_{\lambda}} f \end{bmatrix}, \]
where we are not interested in the second and third row of $\tilde{T}^3+1$.
From the first component,
we have $(1+e^{\pm 2i \Theta_{\lambda}}\lambda - e^{\pm i \Theta_{\lambda}} \lambda) f = 0$.
If $\lambda = 0$, then $f = 0$.
If $\lambda \neq 0$,
we assume that $1+e^{\pm 2i \Theta_{\lambda}}\lambda - e^{\pm i \Theta_{\lambda}} \lambda = 0$.
Then $e^{\pm 2i \Theta_{\lambda}} - e^{\pm i \Theta_{\lambda}} = -\frac{1}{\lambda} \in \MB{R}$,
so we have $\sin (\pm 2\Theta_{\lambda}) - \sin (\pm \Theta_{\lambda}) = 0$.
This holds if and only if $\sin \Theta_{\lambda} (2\cos \Theta_{\lambda} - 1) = 0$.
However, this does not hold because of $\frac{\pi}{3} < \Theta_{\lambda} < \pi$.
Thus, we have $1+e^{\pm 2i \Theta_{\lambda}}\lambda - e^{\pm i \Theta_{\lambda}} \lambda \neq 0$,
so $f = 0$.

(ii) Let $f\in\mathcal{K}$.
By Lemma~\ref{LkerT},
$[f,f,-f]^{\top} \in \ker L = \ker (\tilde{T}^3 + 1)$,
so
\[ \begin{bmatrix} 0 \\ 0 \\ 0 \end{bmatrix}
= (\tilde{T}^3 + 1) \begin{bmatrix} f \\ f \\ -f \end{bmatrix}
= \MMM{2}{2T}{2T}{*}{*}{*}{*}{*}{*} \begin{bmatrix} f \\ f \\ -f \end{bmatrix}
= \begin{bmatrix} 2f \\ * \\ * \end{bmatrix}. \]
From the first component, we have $f = 0$.

(iii) Let $f\in\ker(T+\frac{1}{2})$. 
By Lemma~\ref{LkerT},
$[-f,f,0]^{\top} \in \ker L = \ker (\tilde{T}^3 + 1)$,
so
\[ \begin{bmatrix} 0 \\ 0 \\ 0 \end{bmatrix}
= (\tilde{T}^3 + 1) \begin{bmatrix} -f \\ f \\ 0 \end{bmatrix}
= \MMM{2}{2T}{2T}{*}{*}{*}{*}{*}{*} \begin{bmatrix} -f \\ f \\ 0 \end{bmatrix}
= \begin{bmatrix} -3f \\ * \\ * \end{bmatrix}. \]
From the first component, we have $f = 0$.
\end{proof}


\subsection{The inherited eigenspace}
Now we investigate the inherited eigenspace.
For readers who skipped reading, we briefly review several symbols again.
The matrix $U_c$ is the time evolution of our quantum walk,
which defined by (\ref{K5000}) in Section~3.
The matrix $T$ is the discriminant operator of the Grover walk,
which is defined by (\ref{K5001}) in Section~2.
The subspace $\MC{L}$ is the image of the linear operator $L$,
where $L$ is defined in the beginning of Section~5.
The spectrum of $U_c$ restricted to $\MC{L}$ and the eigenspaces are as follows:

\begin{thm} \label{main thm}
Let $U_{c}, T$ and $\mathcal{L}$ be defined as in above, and let $\{e_u\}_{u\in V}$ be the canonical basis in $\mathbb{C}^V$. 
Then we have
\[ \sigma(U_{c}|_{\mathcal{L}})=\{ e^{\pm i\Theta_{\lambda}} \mid \lambda \in \sigma(T)\backslash\{ 1 \}\} \cup \{ 1\}^{|V|},  \]
where $\Theta_{\lambda}=\cos^{-1}\left(\lambda-\frac{1}{2} \right)$.
The eigenspaces are
\begin{equation} \label{func 1}
\ker(U_{c}-e^{\pm i \Theta_{\lambda}}) \cap \MC{L} = \{(d^{*}+e^{\pm 2i\Theta_{\lambda}}S_{c}d^{*}-e^{\pm i\Theta_{\lambda}}S^{2}_{c}d^{*})f \mid f \in \ker{(T - \lambda )}\}
\end{equation}
whose dimension is $\dim \ker(T-\lambda)$
for $\lambda \in \sigma(T) \setminus \{1, -\frac{1}{2}\}$,
\begin{equation} \label{func 2}
\ker(U_{c} -1) \cap \MC{L} = \{ (d^{*}+S_{c}d^{*}-S^{2}_{c}d^{*}) e_{u} \mid u \in V\}
\end{equation}
whose dimension is $|V|$, and
\begin{equation} \label{func 3}
\ker(U_{c} + 1) \cap \MC{L} =
\left\{ -d^{*}\rho+S_{c}d^{*}\rho \mid \rho \in \ker{ \left(T+\frac{1}{2} \right)} \right\} 
\end{equation}
whose dimension is $\dim \ker(T + \frac{1}{2})$, respectively.
\end{thm}

\begin{proof}
First, we consider the case of $\Lambda \not\in \{1, -1, -\omega, -\omega^2 \}$.
In this case,
\begin{align*}
\ker (U_c - \Lambda) \cap \MC{L}
&= L( \ker(\tilde{T}^3 + 1)(\tilde{T} - \Lambda)) \tag{by (\ref{K212})} \\
&= L( \ker(\tilde{T}+1) \oplus \ker(\tilde{T}+\omega) \oplus \ker(\tilde{T}+\omega^2) \oplus \ker(\tilde{T}-\Lambda))
\tag{by Corollary~\ref{K402}} \\
&= L(\ker(\tilde{T}-\Lambda)) \tag{by Lemma~\ref{LkerT}} \\
&= L \left\{ \begin{bmatrix} f \\ \Lambda^{2}f \\ -\Lambda f \end{bmatrix} \in \MC{K}^3 \; \middle| \;
f \in \ker{\left(T-\frac{\Lambda + \Lambda^{-1} + 1}{2} \right)} \right\}. \tag{by Lemma~\ref{K301}}
\end{align*}
Now, we put $\frac{\Lambda + \Lambda^{-1} + 1}{2} = \lambda$.
Then
\[ \Lambda = \left( \lambda - \frac{1}{2} \right) \pm i \sqrt{ 1 - \left( \lambda - \frac{1}{2} \right)^2 }. \]
Remark that $1 - \left( \lambda - \frac{1}{2} \right)^2 \geq 0$ by Lemma~\ref{K60}.
Since $\Lambda \not\in \{-1, -\omega, -\omega^2 \}$,
we have $\lambda \not\in \{1, \frac{1}{2} \}$.
Put $\Theta_{\lambda} = \cos^{-1} ( \lambda - \frac{1}{2} )$,
and we have $\Lambda = e^{\pm i \Theta_{\lambda}}$.
Thus,
\begin{align*}
\ker(U_c - e^{\pm i \Theta_{\lambda}}) \cap \MC{L}
&= L \left\{ \begin{bmatrix} f \\ e^{\pm 2i \Theta_{\lambda}}f \\ -e^{\pm i \Theta_{\lambda}} f \end{bmatrix} \in \MC{K}^3 \; \middle| \;
f \in \ker{\left(T- \lambda \right)} \right\} \\
&= \{(d^{*}+e^{\pm 2i\Theta_{\lambda}}S_{c}d^{*}-e^{\pm i\Theta_{\lambda}}S^{2}_{c}d^{*})f \mid f \in \ker(T - \lambda ) \}
\end{align*}
is obtained.
As for the dimension, by Lemma~\ref{K411}~(i),
\begin{align*}
\dim \ker(U_c - e^{i \Theta_{\lambda}}) \cap \MC{L}
&= \dim L \left\{ \begin{bmatrix} f \\ e^{2i \Theta_{\lambda}}f \\ -e^{i \Theta_{\lambda}} f \end{bmatrix} \in \MC{K}^3 \; \middle| \;
f \in \ker{\left(T- \lambda \right)} \right\} \\
&= \dim \left\{ \begin{bmatrix} f \\ e^{2i \Theta_{\lambda}}f \\ -e^{i \Theta_{\lambda}} f \end{bmatrix} \in \MC{K}^3 \; \middle| \;
f \in \ker \left(T- \lambda \right) \right\} \\
&= \dim \ker \left(T- \lambda \right).
\end{align*}
Similarly, $\dim \ker(U_c - e^{-i \Theta_{\lambda}}) \cap \MC{L} = \dim \ker \left(T- \lambda \right)$ holds.


Next, we consider the case of $\Lambda = 1$.
By the same argument as in the above case, we have
\begin{align*}
\ker (U_c - 1) \cap \MC{L}
&= L(\ker(\tilde{T}-1)) \\
&= L \left\{ [f,f,-f]^{\top} \in \MC{K}^3 \; \middle| \;
f \in \MC{K} \right\} \tag{by Lemma~\ref{K301}} \\
&= \{ (d^{*}+S_{c}d^{*}-S^{2}_{c}d^{*})f \mid f \in \MC{K}\}.
\end{align*}
Also for this eigenspace, we have
\begin{align*}
\dim \ker (U_c - 1) \cap \MC{L}
&= \dim L \left\{ [f,f,-f]^{\top} \in \MC{K}^3 \; \middle| \;
f \in \MC{K} \right\} \\
&= \dim \left\{ [f,f,-f]^{\top} \in \MC{K}^3 \; \middle| \;
f \in \MC{K} \right\} \tag{by Lemma~\ref{K411}~(ii)}\\
&= |V|.
\end{align*}
In particular,
\[ \ker (U_c - 1) \cap \MC{L} = \{ (d^{*}+S_{c}d^{*}-S^{2}_{c}d^{*}) e_{u} \mid u \in V \}. \]

Finally, we consider the case of $\Lambda \in \{-1, -\omega, -\omega^2 \}$.
Let $j \in \{0,1,2\}$.
We have
\begin{align*}
\ker (U_c + \omega^j) \cap \MC{L}
&= L( \ker(\tilde{T} + \omega^j)^2 \oplus \ker(\tilde{T}+\omega^{j+1}) \oplus \ker(\tilde{T}+\omega^{j+2}) \tag{by Corollary~\ref{K402}} \\
&= L(\ker(\tilde{T} + \omega^j)^2). \tag{by Lemma~\ref{LkerT}}
\end{align*}
If $j \in \{1,2\}$,
Lemma~\ref{K403} derives $L(\ker(\tilde{T} + \omega^j)^2) = L(\ker(\tilde{T} + \omega^j)) = \{ 0 \}$.
If $j = 0$,
we have
\begin{align*}
L(\ker(\tilde{T} + 1)^2)
&= L\left( \ker(\tilde{T} + 1) \oplus
\left\{ [-\rho, \rho, 0]^{\top} \in \MC{K}^3 \; \middle| \; \rho \in \ker \left( T+\frac{1}{2} \right) \right\} \right) \tag{by Lemma~\ref{K404}} \\
&= L \left\{ [-\rho, \rho, 0]^{\top} \in \MC{K}^3 \; \middle| \; \rho \in \ker \left( T+\frac{1}{2} \right) \right\} \\
&= \left\{ -d^{*}\rho+S_{c}d^{*}\rho \; \middle| \; \rho \in \ker \left( T+\frac{1}{2} \right) \right\}.
\end{align*}
Also for this eigenspace, we have
\begin{align*}
\dim \ker (U_c + 1) \cap \MC{L}
&= \dim L \left\{ [-\rho, \rho, 0]^{\top} \in \MC{K}^3 \; \middle| \; \rho \in \ker \left( T+\frac{1}{2} \right) \right\} \\
&= \dim \left\{ [-\rho, \rho, 0]^{\top} \in \MC{K}^3 \; \middle| \; \rho \in \ker \left( T+\frac{1}{2} \right) \right\} \tag{by Lemma~\ref{K411}~(iii)}\\
&= \dim \ker \left( T+\frac{1}{2} \right),
\end{align*}
which completes the proof.
\end{proof}

We note the dimension of $\MC{L}$.
By Theorem~\ref{main thm},
\begin{align*}
    \dim \MC{L} &= 2 \left( |V| - \dim \ker \left(T + \frac{1}{2} \right) - \dim \ker (T - 1) \right) + \dim \ker \left(T + \frac{1}{2} \right) + |V| \\
    &= 3|V| - \dim \ker \left(T + \frac{1}{2} \right) - 2 \dim \ker (T - 1).
\end{align*}
Since $G$ is connected,
$\dim \ker (T - 1) = 1$ by Perron--Frobenius theorem.
Therefore, we have
\begin{equation} \label{0224-0}
\dim \MC{L} = 3|V| - \dim \ker \left(T + \frac{1}{2} \right) - 2.
\end{equation}

Comparing with the conventional spectral mapping theorem,
our theorem has three features.
The first is the correspondence rule of mapping.
In our theorem,
eigenvalues of $T - 1/2$ are lifted up onto the unit circle on the complex plane.
The second is that the eigenspace of $U_c$ that comes from $1 \in \sigma(T)$ vanishes.
The third is, on the other hand,
that the eigenspace of $1$ of $U_c$ are newly born in the inherited eigenspace.

\section{The birth eigenspace} \label{K114}

In this section, we consider the birth eigenspace of $U_c$.
We reveal all eigenvalues come from $\MC{L}^{\perp}$ and their multiplicities.
First, we claim that only $-1, -\omega$ and $-\omega^2$ are eigenvalues come from $\MC{L}^{\perp}$. 

\begin{lem} \label{K131}
{\it
Let $\omega=e^{\frac{2\pi}{3}i}$.
For $\lambda \in \MB{C}$, we have
\[ \ker(U_c - \lambda) \cap \MC{L}^{\perp} = \begin{cases}
\ker d \cap \ker(S_c + \lambda) \quad &\text{if $\lambda \in \{ -1, -\omega, -\omega^2 \}$}, \\
\{ 0 \} \quad & \text{otherwise.}
\end{cases} \]
In particular, 
eigenvalues come from $\MC{L}^{\perp}$ is either $-1, -\omega$ or $-\omega^2$.
}
\end{lem}

\begin{proof}
Recall that $\MC{L}^{\perp} = \ker d \cap \ker dS_c \cap \ker dS_c^2$.
Suppose $\Psi \in \ker(U_c - \lambda) \cap \MC{L}^{\perp}$.
Since $\Psi \in \MC{L}^{\perp} \subset \ker d$,
we have
\[ \lambda \Psi = U_c \Psi = S_c(2d^*d - I)\Psi = S_c2d^*d \Psi - S_c \Psi = - S_c \Psi. \]
Thus, $\Psi$ is an eigenvector of $-S_c$ associated to $\lambda$.
Since $S_c^3 = I$, eigenvalues of $-S_c$ are only $-1$, $-\omega$ and $-\omega^2$.
Conversely, we suppose $\Psi \in \ker d \cap \ker(S_c + \lambda)$ for $\lambda \in \{ -1, -\omega, -\omega^2 \}$.
Then we can check $U_c \Psi = \lambda \Psi$.
For $j \in \{1,2\}$,
we have
\[ dS_c^j \Psi = d(-\lambda)^j \Psi = (-\lambda)^j d\Psi = 0, \]
so $\Psi \in \ker d \cap \ker dS_c \cap \ker dS_c^2 = \MC{L}^{\perp}$.
Thus, we have $\Psi \in \ker(U_c - \lambda) \cap \MC{L}^{\perp}$.
\end{proof}

Define $\MC{B}_{\lambda} = \ker d \cap \ker(S_c + \lambda)$.
This is nothing but the birth eigenspace associated to $\lambda \in \{ -1, -\omega, -\omega^2 \}$.
We supplement some remarks.
The condition $\Psi \in \ker d$ holds if and only if
\begin{equation} \label{K91}
\sum_{a \in \MC{A}(x)} \Psi_a = 0
\end{equation}
for any $x \in V$.

In addition,
the condition $\Psi \in \ker (S_c + \lambda)$ holds if and only if
\[ \Psi_{\tau^{-1}(a)} = -\lambda \Psi_a \]
for any $a \in \MC{A}$.
In particular, when $\lambda = -1$,
we have 
\begin{equation} \label{0219-1}
\Psi_a = \Psi_{\tau(a)} = \Psi_{\tau^{-1}(a)}
\end{equation}
for any $a \in \MC{A}$.

Second, we consider the subspace $\MC{B}_{-1}$.
We use the matrix $R$ defined in (\ref{K200}) to state the following lemma.

\begin{lem} \label{K80}
{\it
The eigenspace $\MC{B}_{-1}$ is isomorphic to $\ker R$.
}
\end{lem}

\begin{proof}
We show that there exists an isomorphism between $\MC{B}_{-1}$ and $\ker R$.
First,
we would like to define the mapping $P : \MC{B}_{-1} \to \ker R$ by
\[ P(\Psi)_{C} = \Psi_a \]
for $\Psi \in \MC{B}_{-1}$, an arc $a$ and a directed triangle $C$ with $a \in C$.
To confirm the well-definedness,
we check $\Psi_a = \Psi_{\tau(a)} = \Psi_{\tau^{-1}(a)}$ and
$\image P \subset \ker R$.
Since $\Psi \in \MC{B}_{-1}$ and (\ref{0219-1}), 
we have $\Psi_a = \Psi_{\tau(a)} = \Psi_{\tau^{-1}(a)}$.
For $\Psi \in \MC{B}_{-1}$ and $x \in V$,
\begin{align*}
(R(P(\Psi)))_x &= \sum_{C \in \pi}R_{x,C} P(\Psi)_C \\
&= \sum_{C \in \pi(x)} P(\Psi)_C \\
&= \sum_{C \in \MC{A}(x)}\Psi_a \tag{by Lemma~\ref{K70}} \\
&= 0 \tag{since $\Psi \in \MC{B}_{-1} \subset \ker d$}
\end{align*}
holds, so $P(\Psi) \in \ker R$.
The mapping $P$ is well-defined.

Next, we would like to define the mapping $Q : \ker R \to \MC{B}_{-1}$ by
\[ Q(\phi)_a = \phi_C \]
for $\phi \in \ker R$, an arc $a$ and a directed triangle $C$ with $a \in C$.
To confirm the well-definedness,
we check $Q(\phi) \in \MC{B}_{-1} = \ker d \cap \ker(S_c - 1)$.
Let $\phi \in \ker R$.
For $x \in V$,
we have
\begin{align*}
(d(Q(\phi)))_x &= \sum_{a \in \MC{A}} d_{x,a} (Q(\phi))_a \\
&= \frac{1}{\sqrt{\deg x}} \sum_{a \in \MC{A}(x)} (Q(\phi))_a \\
&= \frac{1}{\sqrt{\deg x}} \sum_{a \in \pi(x)} \phi_C \tag{by Lemma~\ref{K70}} \\
&= 0, \tag{by $\phi \in \ker R$}
\end{align*}
so $Q(\phi) \in \ker d$.
In addition,
for an arc $a$ and a directed triangle $C$ with $a \in C$,
\begin{align*}
(S_c(Q(\phi)))_a &= \sum_{b \in \MC{A}} (S_c)_{a,b} Q(\phi)_b \\
&= \sum_{b \in \MC{A}} \delta_{a,\tau(b)} Q(\phi)_b \\
&= Q(\phi)_{\tau^{-1}(a)} \\
&= \phi_C \\
&= Q(\phi)_{a}.
\end{align*}
Thus, $S_c(Q(\phi)) = Q(\phi)$, i.e., $Q(\phi) \in \ker(S_c - 1)$.
We have $Q(\phi) \in \ker d \cap \ker(S_c - 1) = \MC{B}_{-1}$.
The mapping $Q$ is well-defined.

Clearly, $P$ and $Q$ is linear mappings.
We finally show that $Q$ is the inverse mapping of $P$.
Let $a$ and $C$ be an arc and a directed triangle $C$ with $a \in C$, respectively.
For $\phi \in \ker R$,
we have
\[ (P \circ Q)(\phi)_C = P(Q(\phi))_C = Q(\phi)_a = \phi_C. \]
For $\Psi \in \MC{B}_{-1}$, we have
\[ (Q \circ P)(\Psi)_a = Q(P(\Psi))_a = P(\Psi)_C = \Psi_a. \]
Therefore,
the linear mapping $P$ is an isomorphism between $\MC{B}_{-1}$ and $\ker R$.
\end{proof}

\begin{cor} \label{K92}
{\it
We have $\dim \MC{B}_{-1} = |\pi| - |V| + b$, where $b = \dim \ker(T + \frac{1}{2})$.
}
\end{cor}

\begin{proof}
Let $D$ be the degree matrix.
We have
\begin{align*}
\dim \MC{B}_{-1} &= \dim \ker R \tag{by Lemma~\ref{K80}} \\
&= |\pi| - \rank R \\
&= |\pi| - \rank R^{\top} \\
&= |\pi| - (|V| - \dim \ker R^{\top}) \\
&= |\pi| - |V| + \dim \ker RR^{\top} \\
&= |\pi| - |V| + \dim \ker (D^{-\frac{1}{2}}R)(D^{-\frac{1}{2}}R)^{\top} \tag{since $D$ is invertible} \\
&= |\pi| - |V| + \dim \ker \left( T + \frac{1}{2} \right) \tag{by (\ref{K90}) in Lemma~\ref{K60}} \\
&= |\pi| - |V| + b,
\end{align*}
which completes the proof.
\end{proof}

Finally, we consider the dimensions of the eigenspaces $\MC{B}_{-\omega}$ and $\MC{B}_{-\omega^2}$.

\begin{lem} \label{K113}
{\it
We have
\[ \dim \MC{B}_{-\omega} = \dim \MC{B}_{-\omega^2} = \frac{1}{2}(|\MC{A}|+2 - 2|V| - |\pi|). \]
}
\end{lem}

\begin{proof}
First, we show that $\dim \MC{B}_{-\omega} = \dim \MC{B}_{-\omega^2}$.
Consider the correspondence $\Psi \mapsto \bar{\Psi}$,
where $\bar{\Psi}$ is the complex conjugate of $\Psi$.
Since $S_c$ is a real matrix, 
this correspondence transfers elements of $\MC{B}_{-\omega}$ and those of $\MC{B}_{-\omega^2}$ to each other.
Also, this correspondence preserves linear independence,
so we have
\[ \dim \MC{B}_{-\omega} \leq \dim \MC{B}_{-\omega^2} \leq \dim \MC{B}_{-\omega}. \]
The dimensions of both eigenspaces are equal.
Next, we find their dimensions.
By (\ref{0224-0}) and Corollaly~\ref{K92},
\begin{align*}
2 \dim \MC{B}_{-\omega} &=  |\MC{A}| - (\dim \MC{L} + \dim \MC{B}_{-1}) \\
&= |\MC{A}| - \{ (3|V| - b - 2 ) + (|\pi| - |V| + b) \} \\
&= |\MC{A}| + 2 - 2|V| - |\pi|,
\end{align*}
where $b = \dim \ker(T + \frac{1}{2})$.
We have the statement.
\end{proof}

Note that $\pi$ is a partition of $\MC{A}$ consisting of directed triangles,
so it holds that $|\MC{A}| = 2|E| = 3|\pi|$.
Thus, Corollary~\ref{K92} and Lemma~\ref{K113} are summarized as follows: \begin{align*}
\dim \MC{B}_{-1} &= \frac{2}{3}|E| - |V| + \dim \ker \left( T + \frac{1}{2} \right), \\
\dim \MC{B}_{-\omega} &= \frac{2}{3}|E| - |V| + 1, \\
\dim \MC{B}_{-\omega^2} &= \frac{2}{3}|E| - |V| + 1.
\end{align*}

\section{Double cones} \label{K150}
In this section, we analyze the spectrum of double cones,
a family of triangulable graphs.
On this example, not only eigenvalues of the time evolution but also eigenvectors are completely revealed.

\begin{figure}[h]
\begin{center}
\begin{tikzpicture}
[line width=1pt, scale = 0.8,
v/.style = {circle, fill = black, inner sep = 1.1mm},
g/.style = {circle, fill = black, inner sep = 0mm}]
\node[v] (1) at (1.9*2.5, 0.62*2/3) {};
\node[v] (2) at (0*2.5, 2*2/3) {};
\node[v] (3) at (-1.9*2.5, 0.62*2/3) {};
\node[v] (4) at (-1.17*2.5, -1.62*2/3) {};
\node[v] (5) at (1.17*2.5, -1.62*2/3) {};
\node[v] (6) at (0, 5) {};
\node[v] (7) at (0, -5) {};
\draw (1) -- (2) -- (3) -- (4) -- (5) -- (1);
\draw (7) -- (1);
\draw (7) -- (2);
\draw (7) -- (3);
\draw (7) -- (4);
\draw (7) -- (5);
\draw (6) -- (1);
\draw (6) -- (2);
\draw (6) -- (3);
\draw (6) -- (4);
\draw (6) -- (5);
\node[g, label =  above : $u_+$] at (6) {};
\node[g, label =  below : $u_-$] at (7) {};
\node[g, label =  right : $x_0$] at (1) {};
\node[g, label =  above left : $x_1$] at (2) {};
\node[g, label =  left : $x_2$] at (3) {};
\node[g, label =  below right : $x_3$] at (-1.17*2.5 + 0.17, -1.62*2/3) {};
\node[g, label =  below left : $x_4$] at (1.17*2.5 - 0.17, -1.62*2/3) {};
\end{tikzpicture}
\caption{The double cone $\Gamma_5$}
\label{K120}
\end{center}
\end{figure}
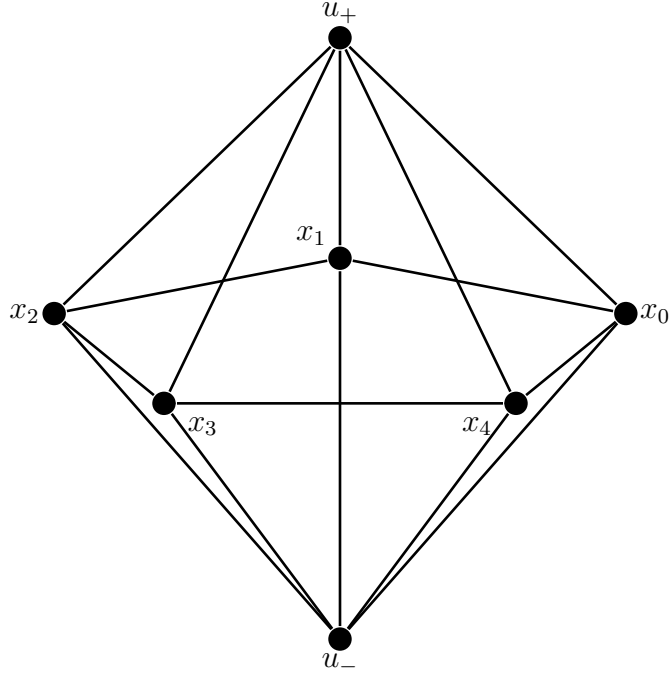

Let $G$ and $H$ be graphs.
The {\it join} $G + H$ of $G$ and $H$ is defined by
\begin{align*}
V(G + H) &= V(G) \cup V(H), \\
E(G + H) &= E(G) \cup E(H) \cup \{ xy \mid x \in V(G), y \in V(H) \}.
\end{align*}
Roughly speaking, this is a graph considered $G$ and $H$ as one,
with all possible edges between $G$ and $H$.
The {\it double cone} $\Gamma_n$ is the graph defined by $\Gamma_n = \overline{K_2} + C_n$,
where $\overline{K_2}$ is the complement of the complete graph $K_2$,
and $C_n$ is the cycle on $n$ vertices.
For convenience, the vertices of $\Gamma_n$ are labeled as
$V(\overline{K_2}) = \{ u_+, u_- \}$, $V(C_n) = \{ x_0, x_1, \cdots, x_{n-1} \}$.
The double cone $\Gamma_5$ is shown in Figure~\ref{K120}.

Giving the partition
\[ \pi = \bigcup_{i=0}^{n-1}\{ (u_+, x_i), (x_i, x_{i+1}), (x_{i+1}, u_+) \} \cup
\bigcup_{i=0}^{n-1}\{ (u_-, x_{i+1}), (x_{i+1}, x_i), (x_{i}, u_-) \}
\]
of $\MC{A}(\Gamma_n)$, we see that $\Gamma_n$ is triangulable,
where we set $x_n = x_0$.
We analyze eigenvalues of the time evolution $U_c$
of the quantum walk determined by $\Gamma_n$ and this $\pi$.

\subsection{The spectrum of the discriminant operator of $\Gamma_n$} \label{K201}
In order to clarify the inherited eigenspace, we analyze the eigenvalues of the discriminant operator $T = T(\Gamma_n)$.
We denote the $m \times n$ all-ones matrix by $J^{(m,n)}$.
By indexing $V(\Gamma_n)$ as $\{u_+, u_-, x_0, \dots, x_{n-1} \}$,
we see that the adjacency matrix $A = A(\Gamma_n)$ of a double cone is
\[ A(\Gamma_n) = \MM{O}{J^{(2,n)}}{J^{(n,2)}}{A(C_n)}. \]
In addition,
the degree matrix is $D = D(\Gamma_n) = \diag (n, n, 4, 4, \dots, 4)$.
Remark that $D^{-\frac{1}{2}} T D^{\frac{1}{2}} = D^{-1}A$,
so $D^{-1}A$ is similar to the discriminant operator $T$.
Thus, we define $T' = D^{-1}A$ and analysis eigenvalues of $T'$ instead of $T$.
This modified discriminant operator $T'$ is written as
\[ T' = \MM{O}{\frac{1}{n} J^{(2,n)}}{\frac{1}{4} J^{(n,2)}}{ \frac{1}{4}A(C_n)}. \]
Define matrices $\tilde{J} \in \MB{C}^{(n+2) \times 2}$ by
\[ \tilde{J} = \MM{J^{(2,1)}}{O}{O}{J^{(n,1)}}, \]
and $Q' \in \MB{C}^{2 \times 2}$ by
\[ Q' = \MM{0}{1}{\frac{1}{2}}{\frac{1}{2}}. \]
Since $A(C_n)J^{(n,1)} = 2J^{(n,1)}$, we have
\begin{equation} \label{K100}
T'\tilde{J}  = \tilde{J}Q'.
\end{equation}

In the discussion below,
we first construct $n + 2$ eigenvectors of $T'$,
and then describe that the constructed eigenvectors are linearly independent.

\begin{lem} \label{K110}
Let $v$ be an eigenvector of $Q'$ associated to an eigenvalue $\lambda$.
Then, $\tilde J v$ is an eigenvector of $T'$ associated to an eigenvalue $\lambda$.
\end{lem}

\begin{proof}
By Equality~(\ref{K100}),
\[ T'(\tilde{J}v) = (T'\tilde{J})v = (\tilde{J}Q')v = \tilde{J}(Q'v) = \tilde{J}(\lambda v) = \lambda (\tilde{J}v), \]
so we have the statement.
\end{proof}

The characteristic polynomial of $Q'$ is $(\lambda - 1)(\lambda + \frac{1}{2})$,
so we have two eigenvectors of $T'$ by Lemma~\ref{K110}.
Next, we construct $n-1$ eigenvectors of $T'$ from eigenvectors of $A(C_n)$.

\begin{lem}[\cite{Spg}, Subsection~1.4.3] \label{K111}
{\it
Let $\zeta = e^{\frac{2\pi i}{n}}$,
and let $v_{j} = [1, \zeta^j, \zeta^{2j}, \dots, \zeta^{(n-1)j}]^{\top} \in \MB{C}^{V(C_n)}$.
Then, $v_{j}$ is an eigenvector of $A(C_n)$ associated to the eigenvalue $\zeta^j + \zeta^{-j}$.
In particular,
\[ \sigma(A(C_n)) = \left\{ 2 \cos \frac{2\pi}{n} j \, \middle | \, j = 0,1, \dots, n-1 \right\}. \]
}
\end{lem}

Let ${\bm 0}_m$ be the zero vector of length $m$.
Remark that $J^{(s,n)}v_j={\bm 0}_s$.

\begin{lem} \label{K112}
{\it
Let $v_{j}$ be the vector defined in Lemma~\ref{K111}.
Define $\tilde{v}_{j} = [{\bm 0}_2^{\top} \; v_{j}^{\top}]^{\top}$.
If $j \neq 0$,
then the vector $\tilde{v}_{j}$ is an eigenvector of $T'$ associated to $\frac{1}{4}(\zeta^j + \zeta^{-j})$.
}
\end{lem}

\begin{proof}
Eigenvectors for different eigenvalues are orthogonal to each other,
so if $j \neq 0$, the vector $v_{j}$ is orthogonal to all-ones vector.
Thus, we have
\[ T' \tilde{v_{j}} = \MM{O}{\frac{1}{n} J^{(2,n)}}{\frac{1}{4} J^{(n, 2)}}{\frac{1}{4} A(C_n)} \bC{{\bm 0}_2}{v_{j}}
= \bC{{\bm 0}_2}{\frac{1}{4}A(C_n) v_{j} }
= \frac{1}{4}(\zeta^j + \zeta^{-j}) \bC{{\bm 0}_2}{v_{j}} = \frac{1}{4}(\zeta^j + \zeta^{-j}) \tilde{v_{j}}, \]
which completes the proof.
\end{proof}

The last eigenvector of $T'$ is $\tilde{v} = [1, -1, {\bm 0}_n^{\top}]^{\top}$.
It is easy to check that $\tilde{v}$ is an eigenvector of $T'$ associated to $0$.
We now have the $n+2$ eigenvectors of $\tilde{T}$.
The argument left is on linear independence.

\begin{prop} \label{K130}
{\it
The spectrum of $T'$ is
\[ \sigma(T') = \sigma(T) = \{ 0 \}
\cup \left\{ \frac{1}{2}\cos \frac{2\pi}{n}j \; \middle| \; j = 1,2, \dots, n-1  \right\}
\cup \left\{ 1, -\frac{1}{2} \right\}. \]
Moreover, their eigenvectors are the vectors $\tilde{v}$, $\tilde{v}_{j}$ in Lemma~\ref{K112},
and $\tilde{J}v$ in Lemma~\ref{K110}, respectively.
}
\end{prop}

\begin{proof}
We check linear independence of the eigenvectors that we obtained.
Since $\tilde{v}$, $\tilde{v}_{j} \in \ker \tilde{J}^* = (\image \tilde{J})^{\perp}$,
the argument of linear independence is divided into the two cases,
where each eigenvectors are in $\image \tilde{J}$ and $(\image \tilde{J})^{\perp}$.
The linear independence of $\tilde{v}$ and $\tilde{v}_{j}$ is clear by difference in their supports.
The linear independence of $\tilde{J}v$, where $v$ is an eigenvector of $Q'$,
can be seen from difference of eigenvalues.
Therefore, the $n+2$ eigenvectors are certainly linearly independent.
\end{proof}

Now that eigenvalues and eigenvectors of $T'$ have been revealed,
the inherited eigenspace of $\Gamma_n$ also has been revealed by Theorem~\ref{main thm}.

\subsection{The birth eigenspace of $\Gamma_n$}

Next, we investigate the birth eigenspace.
Since the number of eigenvectors to be found is known from Section~\ref{K114},
we will explicitly construct eigenvectors by using features of $\Gamma_n$.
By Proposition~\ref{K130}, we have
\[ b = \dim \ker \left( T(\Gamma_n ) + \frac{1}{2} \right) = \begin{cases}
1 \quad &\text{if $n$ is odd,} \\
2 \quad &\text{if $n$ is even,}
\end{cases} \]
so by Corollary~\ref{K92} and Lemma~\ref{K113},
\[ \dim \MC{B}_{-1} 
= \begin{cases}
n-1 \quad &\text{if $n$ is odd,} \\
n \quad &\text{if $n$ is even,}
\end{cases} \]
and
\[ \dim \MC{B}_{-\omega} = \dim \MC{B}_{-\omega^2} 
= n-1. \]
For a vector $\Psi \in \MB{C}^{\MC{A}}$ and indices $j \in \{ 0, 1, \dots, n-1\}$,
we define $a_j = \Psi_{(x_j, x_{j+1})}$ and $b_j = \Psi_{(x_{j+1}, x_{j})}$,
where the number $n$ that appears in the subscript should be read as 0, as appropriate.
Since $\ker(U_c - \lambda) \cap \MC{L}^{\perp} = \ker d \cap \ker(S_c + \lambda)$ by Lemma~\ref{K131},
proper determination of $a_j$ and $b_j$ is key.

\begin{lem} \label{K140}
{\it
Suppose $k \in \{0,1,2\}$ and $\Psi \in \ker(S_c - \omega^k)$.
Then $\Psi \in \ker d$ if and only if both of the following are satisfied:
\begin{enumerate}[(1)]
\item $a_j + \omega^{2k}a_{j+1} + \omega^{2k}b_{j} + b_{j+1} = 0$; and
\item $\sum_{j = 0}^{n-1} a_j = \sum_{j = 0}^{n-1} b_j = 0$.
\end{enumerate}
}
\end{lem}

\begin{proof}
Assume that $\Psi \in \ker d$.
By Equality~(\ref{K91}), we have
\begin{align*}
0 &= \sum_{a \in \MC{A}(x_{j+1})} \Psi_a \\
&= a_j + \Psi_{(u_+, x_{j+1})} + b_{j+1} + \Psi_{(u_-, x_{j+1})} \\
&= a_j + \omega^{2k}a_{j+1} + \omega^{2k}b_{j} + b_{j+1}. \tag{by $\Psi \in \ker(S_c - \omega^k)$}
\end{align*}
In addition,
\[ 0 = \sum_{a \in \MC{A}(u_+)} \Psi_a = \sum_{j=0}^{n-1} \Psi_{(x_j, u_+)} = \omega^k \sum_{j=0}^{n-1} a_j. \]
Similarly, we have $\sum_{j=0}^{n-1} b_j = 0$.
The opposite follows the same calculation from the bottom.
\end{proof}

In constructing vectors $\Psi$ in $\MC{B}_{-\omega^k}$,
we need only declare values of $a_j$ and $b_j$.
This is because the other components of $\Psi$ are determined as belonging to $\ker (S_c - \omega^k)$.
Let $\zeta = e^{\frac{2\pi i}{n}}$.
By defining $a_j$ and $b_j$ as shown in Table~\ref{K141} for $l \in \{1,2, \dots, n-1\}$,
we first get $n-1$ vectors in $\MC{B}_{-\omega^k}$.
Indeed, 
it is easy to confirm that the two conditions of Lemma~\ref{K140} are satisfied.

\begin{table}[h]
  \begin{center}
    \begin{tabular}{c|c|c} \hline
    Eigenspace & $a_j$ & $b_j$ \\ \hline
    $\MC{B}_{-1}$ & $\zeta^{lj}$ & $- \zeta^{lj}$ \\
    $\MC{B}_{-\omega}$ & $-(\omega^2 + \zeta^l)\zeta^{lj}$ & $(1+\omega^2 \zeta^l) \zeta^{lj}$  \\ 
    $\MC{B}_{-\omega^2}$ & $-(\omega + \zeta^{-l})\zeta^{-lj}$ & $(1+\omega \zeta^{-l}) \zeta_n^{-lj}$ \\ \hline    
    \end{tabular}
    \caption{Vectors in $\MC{B}_{-\omega^k}$}
    \label{K141}
  \end{center}
\end{table}

When $n$ is even, consider another vector defined by $a_j = b_j = (-1)^j$
in addition to the vectors in Table~\ref{K141}.
This vector also belongs to $\MC{B}_{-1}$.
For each eigenspace,
it can be confirmed by calculation that the constructed vectors are orthogonal to each other.
Therefore, the vectors we have are linearly independent.
Now, we have got all explicit eigenvectors of the time evolution $U_c$
of the double cone $\Gamma_n$.

\section{Summary and future works}

In this paper,
we define a new quantum walk for triangulable graphs
and clarify the spectral mapping theorem between the time evolution and the discriminant operator.
Its corresponding rule is to translate eigenvalues of the discriminant operator by $-1/2$
and then lift up them on the unit circle on the complex plane.
As for the birth eigenspace,
the multiplicities of eigenvalues are completely determined.
Eigenvectors of the birth eigenspace in the case of the Grover walks
can be explicitly constructed by using fundamental cycles,
while in the case of our quantum walk, nothing corresponding to them has been found,
so, at present, we have not yet explicitly constructed eigenvectors of the birth eigenspace.
Thus, in Section~\ref{K150},
we have tried to explicitly construct eigenvectors of the birth eigenspace for the double cones as concrete graphs,
but these seem to strongly depend on the shape of $\Gamma_n$,
so general construction will be a future work.
Also, it is not clear which property of graphs determine the multiplicity of the minimum eigenvalue of the discriminant operator.
In the case of the Grover walks,
its multiplicity can be determined by whether graphs are bipartite or not.
In the future, it should be clarified which property of graphs determines the multiplicity of the smallest eigenvalues.



\end{document}